\documentclass[aps,superscriptaddress]{revtex4}
\usepackage{amssymb,amsmath,epsfig}
\usepackage[colorlinks=true, pdfstartview=FitV, linkcolor=blue, citecolor=red, urlcolor=magenta, breaklinks=true]{hyperref}
\begin{document}
\title{ Superresonance Phenomenon from Acoustic Black Holes in Neo-Newtonian theory}
\author{I. G. Salako}
\email{ ines.salako@imsp-uac.org; inessalako@gmail.com}
\affiliation{Institut de Math\'ematiques et de Sciences Physiques
(IMSP) 01 BP 613 Porto-Novo, B\'enin}
\author{Abdul Jawad}
\email{jawadab181@yahoo.com; abduljawad@ciitlahore.edu.pk}
\affiliation{Department of Mathematics, COMSATS Institute of
Information Technology, Lahore-54000,\\ Pakistan}

\begin{abstract}

We explore the possibility of the acoustic analogue of a
super-radiance like phenomenon, i.e., the amplification of a sound
wave by reflection from the ergo-region of a rotating acoustic black
hole in the fluid draining bathtub model in the presence of the
pressure be amplified or reduced in agreement with the value of the
parameter $\left(\gamma=1+ \frac{kn\rho_0^{n-1} }{c^2}\right)$. We
remark that the interval of frequencies depend upon the
neo-newtonian parameter $\gamma$\, ($\bar{\Omega}_H = \frac{2
}{1+\gamma}\;\Omega_{H}$)\textbf{ and becomes narrow in this work}.
As a consequence, the tuning of the neo-newtonian parameter
$\left(\gamma=1+ \frac{kn\rho_0^{n-1} }{c^2}\right)$ changes the
rate of loss of the acoustic black hole mass.

\end{abstract}
\maketitle
\pretolerance10000

\section{Introduction}

Acoustic black holes possess many of the fundamental properties of
the black holes of general relativity and have been extensively
studied in the literature \cite{M,Volovik,fcreation}. The connection
between black hole physics and the theory of supersonic acoustic
flow was established in 1981 by Unruh \cite{fcreation} and has been
developed to investigate the Hawking radiation and other phenomena
for understanding quantum gravity. Hawking radiation is an important
quantum effect of black hole physics. In 1974, Hawking combining
Einstein's General Relativity and Quantum Mechanics announced that
classically a black hole does not radiate, but when we consider
quantum effects emits thermal radiation at a temperature
proportional to the horizon surface gravity.

Since the Hawking radiation showed by Unruh \cite{fcreation} is a
purely kinematic effect of quantum field theory, we can study the
Hawking radiation process in completely different physical systems.
For example, acoustic horizons are regions where a moving fluid
exceeds the local sound speed through a spherical surface and
possesses many of the properties associated with the event horizons
of general relativity. In particular, the acoustic Hawking radiation
when quantized appears as a flux of thermal phonons emitted from the
horizon at a temperature proportional to its surface gravity. Many
fluid systems have been investigated on a variety of analog models
of acoustic black holes, including gravity wave~\cite{RS}, water
\cite{Mathis}, slow light \cite{UL}, optical fiber \cite{Philbin}
and  electromagnetic waveguide \cite{RSch}. The models of superfluid
helium II \cite{Novello}, atomic Bose-Einstein condensates
\cite{Garay,OL} and one-dimensional Fermi degenerate noninteracting
gas \cite{SG} have been proposed to create an acoustic black hole
geometry in the laboratory.

%
%
The analogous systems employ a classical as well as Newtonian
treatment generally, and also some quantum systems are considered.
Also, a BH is a relativistic gravitational phenomenon which requires
the inertial and gravitational effects of the pressure for a
reasonable description of the system. However, relativistic pressure
effects can be incorporated in a Newtonian framework in some cases
approximatively. This is called neo-Newtonian theory and it is
modification of the usual Newtonian theory by comprising the enough
pressure into the dynamics.

In the present work, our main goals is firstly to realize the
process of drawing the acoustic BH for the neo-Newtonian
hydrodynamics, on the other hand, to analyze the impact of
neo-Newtonian parameter on the superresonance  phenomenon especially
on the frequencies of the waves. The paper is organized as follows.
In Sec. \ref{neonewtonian} we will review the development process of
the acoustic BH after recalling the basic concepts of the
neo-Newtonian theory \cite{ines}. In Sec. \ref{superresonance} we
will develop superresonance phenomenon and in Sec. \ref{numerical}
the numerical results.  In Sec. \ref{conclusion} we make our final
conclusions.

\section{Newtonian Hydrodynamics in an Expanding Background: Cosmology} \label{neonewtonian}

In this section, we present the neo-Newtonian hydrodynamics applied
to cosmology. However, we consider the standard case of Newtonian
equations firstly. In the presence of inviscid perfect fluid, the
basic equations of Newtonian hydrodynamics are
\begin{eqnarray}
\dot{\rho} + \nabla\cdot(\rho\vec{v})  &=& 0, \label{eq100'}\\
\rho \frac{d \vec{v}}{dt} \equiv  \rho [\dot{\vec{v}}+
(\vec{v}\cdot\nabla)\vec{v} ] &=& -\nabla p, \label{eq200'}
\end{eqnarray}
where $\rho,~p$ and $\vec v$ are fluid density, corresponding
pressure and the velocity field, respectively. The dot shows the
differentiation with respect to cosmic time $t$. The above system of
equations becomes suitable to study cosmology adopting the velocity
field $\vec{v}=H (t) \vec{r}$ (Hubble's law) where
$H(t)=\frac{\dot{a}(t)}{a(t)}$, being $a(t)$ the scale factor. It is
worth noting the trivial solution for the continuity Eq.
(\ref{eq100'}) $\rho(a)=\rho_0/a^3$, where the today's scale factor
$a_0=1$ gives the today's density of the fluid $\rho_0$.

Gravitational interaction is coupled into Euler's equation (\ref{eq200'}) as
\begin{equation}\label{euler1} \dot{\vec{v}}+
(\vec{v}\cdot\nabla)\vec{v}  = - \frac{ \nabla p}{\rho} - \nabla \Psi ,
\end{equation}
where the gravitational potential $\Psi$ obeys the Poisson equation
\begin{equation}
\label{poisson1}\nabla^{2} \Psi = 4 \pi G \rho.
\end{equation}
Eqs. (\ref{eq100'}) and (\ref{euler1}) represents the fluid picture
of cosmic medium which is gravitationally self-interacting through
Poisson Eq. (\ref{poisson1}). In the framework of Newtonian
cosmology, the Friedmann equations are given as follows
\begin{equation}
\frac{\dot{a}^2}{a^2}+\frac{\left(-2E\right)}{a^2}=\frac{8\pi G}{3}
\rho \hspace{0.5cm}{\rm and}\hspace{0.5cm}\dot{H}+H^2=-\frac{4\pi
G}{3}\rho,
\end{equation}
where $E$ appears as integration constant associated to the energy
of system. Moreover, the pressure does not correspond to homogeneous
and isotropic background. With the inclusion of Newtonian cosmology,
we can not model a radiation dominated as well as dark energy
dominated epoch. This approach is restricted to a description of the
Einstein-de Sitter universe.

\subsection{Including Pressure: The neo-Newtonian Theory}

The neo-Newtonian equations has been developed by McCrea \cite{C}
and Harrison \cite{m2} which ensures the effects of pressure as well
the simplicity of Newtonian physics. Later, a crucial study related
to the perturbative behavior of the neo-Newtonian equations has
helped in developing the final expression for the fluid equations in
this approach \cite{Lima1997} (see also \cite{RRRR2003, velten}) which are
given by
\begin{equation}
\partial_{t} \rho_i + \nabla\cdot(\rho_i\vec v) + p \nabla\cdot\vec{v} = 0\;.
\label{salako1bis}
\end{equation}
\begin{equation}\label{salako2}
\dot{ \vec{v}} + (\vec{v} \cdot  \nabla )\vec{v} = - \nabla \Psi   - \frac{ \nabla  p}{\rho + p}\,,
\end{equation}
\begin{equation}
\label{salako3}\nabla^{2} \Psi = 4 \pi G \left(\rho + 3 p\right).
\end{equation}
As is widely known, the above set of equations admits a homogeneous
and isotropic solution, i.e. , $p=p(t)$ and $\rho=\rho(t)$. In this
case, the fluid velocity is (a dot means partial time derivative)
\begin{equation}
\vec{v}= {\dot{a}\over a}\vec{r}\,  ,\label{eq:4}\\
\end{equation}

Combining Eqs.~(\ref{eq:4}), (\ref{salako2}) and (\ref{salako3}), we can
obtain the following equations
\begin{eqnarray}
\frac{\dot{a}^2}{a^2}+\frac{\left(-2E\right)}{a^2}=\frac{8\pi G}{3} \rho ,\\
\dot{H}+H^2=-\frac{4\pi G}{3}(\rho+3p).
\end{eqnarray}
with the continuity equation (\ref{salako1bis}) reducing to the following form
\begin{equation}
{\partial \rho \over \partial t }+ 3{\dot{a}\over a}(\rho + {p }) = 0\, .
\label{eq:7}
\end{equation}
These equations are exactly correspond to the relativistic Friedmann
equations. The main idea behind the neo-Newtonian formalism relies
on the following substitutions: Firstly, it is necessary to redefine
the concept of inertial and passive gravitational mass density. With
the redefinition
\begin{equation}\label{rhoi}
\rho_i \rightarrow \rho + p,
\end{equation}
we rewrite the continuity and the Euler equation.

The second step is the interpretation of the active gravitational mass density i.e., the density that source
the gravitational field. Hence the following redefinition
\begin{equation}\label{rhog}
\rho_g \rightarrow \rho + 3p,
\end{equation}
will become the source of the Poisson equation.
The generalization of this result in the presence of pressure has
been evaluated in \cite{C}. Moreover, this approach has been
modified \cite{m2} which leads to neo-Newtonian theory.

When $p = 0$ we find the Newtonian equations. The interesting
feature of above equations is that the inertial mass present in the
Newtonian equation can be replaced by $\rho + \frac{p}{c^2}$. From
cosmological structure formation point of view, it is noted that
equation (\ref{salako1bis}) does not provide correct growth of
matter for large scale perturbations in the scenario of homogeneous,
isotropic and expanding background \cite{ademir,rrrr,velten}.
Actually, the correct growth for large scale perturbations
 is obtained  if Eq.(\ref{salako1bis}) is modified as \cite{ademir,rrrr,velten}
\begin{eqnarray}
\partial_{t} \rho + \nabla\cdot(\rho\vec v) + p \nabla\cdot\vec{v} &=& 0\;. \label{eq1bis}
\end{eqnarray}
This equation shows correspondence with effective metric of usual
Newtonian scenario when applied to fluid configuration considered.
It represents that one can see, how the neo-Newtonian formulation is
sensible to the the specific symmetries and hypothesis of the
problem in another context of the cosmological one. This fact may
indicate that the construction of a Newtonian counterpart of a
relativistic problem may vary from problem to problem, deserving in
the general case a deeper analysis.

\subsection{Acoustic Black Holes in neo-Newtonian Theory}

Let us now consider the barotropic fluid, i.e. $ p=p(\rho)$,
inviscid and irrotational where the equation of state $p = k
\rho^{n}$, with $k$ and $n$ constants. We write the fluid velocity
as $ \vec{v}=-\nabla\psi $ where $ \psi $ is the velocity potential.
Thus, we linearise the equations (\ref{salako1bis}), (\ref{salako2})
and (\ref{eq1bis}) by perturbing $ \rho $, $ \vec{v} $ and $ \psi $
as follows:
\begin{eqnarray}
 \rho &=& \rho_{0} + \varepsilon \rho_{1} + 0(\varepsilon^{2})\;, \\
 \rho^{n} &=& \left[\rho_{0} + \varepsilon \rho_{1}+ 0(\varepsilon^{2})\right]^{n} \approx
 \rho^{n}_{0} + n \varepsilon \rho^{n-1}_{0} \rho_{1} + ... \;,
 \\
 \vec{v} &=& \vec{v}_{0} + \varepsilon \vec{v}_{1} + 0(\varepsilon^{2}),
 \\
\psi &=& \psi_{0} + \varepsilon\psi+ 0(\varepsilon^{2})\;, \label{salako80'}
 \end{eqnarray}
 where $\rho$ is the fluid density, $p$ its pressure and $\vec v$ its the velocity field.
Thus, the wave equation becomes
\begin{eqnarray}
&-&\partial_{t} \Big\{c_{s}^{-2}\rho_{0}\Big[\partial_{t}\psi +
\Big(\frac{1}{2} + \frac{\gamma}{2}\Big)\vec v_{0}.\nabla
\psi\Big]\Big\} +\nabla\cdot\Big\{- c_{s}^{-2}\rho_{0}
\vec{v_{0}}\Big[\Big(\frac{1}{2} +
\frac{\gamma}{2}\Big)\partial_{t}\psi
\gamma\vec{v_{0}}.\nabla \psi\Big]+ \rho_{0}\nabla
\psi\Big\}=0,
\label{salako12bis}
\end{eqnarray}
and we can rewrite as follows
\begin{eqnarray}
\partial_{\mu} (f^{\mu\nu} \partial_{\nu} \psi) = 0
\end{eqnarray}
where
\begin{equation}
{  f^{\mu\nu}(t,\vec{x}}) = \frac{\rho_{0}}{ c^2_{s}} \left(
\begin{array}{ccc}
-1  & - \frac{1+\gamma}{2}v^x & - \frac{1+\gamma}{2}v^y \\
- \frac{1+\gamma}{2}v^x  &    c^2_s - (v^x)^2           & -\gamma \;v^x\;v^y \\
- \frac{1+\gamma}{2}v^y&  -\gamma \;v^x\;v^y & c^2_s - (v^x)^2
\end{array}
\right).
\end{equation}
Here we are assuming a (2+1)-dimensional spacetime.
However, considering the Klein-Gordon equation \cite{velten} for a massless scalar field we obtain
\begin{eqnarray}
 \frac{1}{\sqrt{-g}} \partial_{\mu} ( \sqrt{-g} g^{\mu\nu} \partial_{\nu} \psi) = 0,   \label{salako14'}
\end{eqnarray}
and so the effective (acoustic) metric, in cylindrical  coordinates, reads
\begin{eqnarray}\label{salako15''}
ds^{2} &=& \sqrt{\frac{\rho^2_0}{c^2_s+ (v^2_r +
v^2_{\phi})(\frac{\gamma -1}{2})^2 }} \Big [- c^2_s \; dt^2  +
(dr-\gamma \; v_r \;dt)^2 (rd\phi+ \gamma \;v_\phi \;dt)+dz^2\Big],
\end{eqnarray}
where $ \gamma=1+ \frac{kn\rho_0^{n-1} }{c^2}  $.
\subsection{Ergo-region and event horizon}

Now considering a static and position independent density, the velocity field is given by

\begin{eqnarray}
\vec{v} = \frac{A}{r} \hat r + \frac{B}{r} \hat\phi,
\label{salako17''}
\end{eqnarray}
which is a solution obtained from the continuity equation (\ref{salako1bis}) and the velocity potential is
\begin{eqnarray}
 \psi(r,\phi) = - A \ln r - B \phi.    \label{salako18'}
\end{eqnarray}
Thus, considering (\ref{salako17''}), (\ref{salako18'}) and also the coordinate transformations as follows
\begin{eqnarray}
dt = dt' + a_1 dr, \quad d\phi = d\phi' + a_2 dr,
\end{eqnarray}
being
\begin{eqnarray}
a_1 = \frac{A \;r\; (1+\gamma )}{2(A^2 \;\gamma-cs^2 \;r^2)}, \quad
a_2 = \frac{A\; B\; \gamma }{r(A^2\; \gamma-cs^2 \;r^2)},
\end{eqnarray}
into the metric (\ref{salako15''}) we obtain the acoustic black hole
in neo-Newtonian theory  which is given by~\cite{ines}
\begin{eqnarray}
\label{m-ab-nn}
ds^2&=&\rho_0\beta_1\left[-\left(1-\frac{r^2_e}{r^2}\right)dt^2+(1+\beta_2)
\left(1-\frac{r_h^2}{r^2}\right)^{-1} dr^2-\frac{2B\beta_3}{r}rdtd\phi+
\Big(1+\frac{\beta_4}{r^2} \Big) r^2d\phi^2\right].
\end{eqnarray}
where
\begin{eqnarray}
\beta_1 &=&\left(1+\beta_2\right)^{-1/2},
\quad\quad
\beta_2 =\frac{r^2_e}{r^2}\left(\frac{\gamma-1}{2}\right)^2,
\\
\beta_3 &=&\frac{(1+\gamma)}{2}, \quad\quad \beta_4=\left(\frac{A(\gamma-1)}{2}\right)^2,
\end{eqnarray}
being  $ r_e $ the radius of ergo-region and $r_h$ the event horizon, i.e.,
\begin{eqnarray}
r_e=\sqrt{\gamma(A^2+B^2)},  \quad\quad r_h=\sqrt{\gamma}\vert A\vert.
\end{eqnarray}
Thus, the metric (\ref{m-ab-nn}) can be now written in the form
\begin{eqnarray}\label{metrinv}
g_{\mu\nu}=\rho_0\beta_1\left[\begin{array}{clcl}
-f &\quad\quad\quad 0& -\frac{B\beta_3}{r}\\
0 & \quad (1+\beta_2){\cal Q}^{-1}& 0\\
-\frac{B\beta_3}{r} &\quad\quad\quad 0 & \left( 1+\frac{\beta_4}{r^2} \right)
\end{array}\right],
\end{eqnarray}
and the inverse of the $g_{\mu\nu}$ is
\begin{eqnarray}
g^{\mu\nu}=\frac{\rho_0\beta_1(1+\beta_2)}{-g}\left[\begin{array}{clcl}
-\Big(1+\frac{\beta_4}{r^2}\Big){\cal Q}^{-1} &\quad\quad 0& \quad\quad-\frac{B\beta_3}{r{\cal Q}}\\
0 & \quad \frac{-g{\cal Q}}{(1+\beta_2)^2}&\quad\quad 0\\
-\frac{B\beta_3}{r{\cal Q}} &\quad\quad 0 & \quad\quad \frac{f}{{\cal Q}}
\end{array}\right],
\end{eqnarray}
where
\begin{eqnarray}
f&=&1-\frac{r_e^2}{r^2}, \quad\quad {\cal Q}=1-\frac{r_h^2}{r^2},
\\
-g&=&\frac{(1+\beta_2)}{{\cal Q} }\left[\left(1+\frac{\beta_4}{r^2}\right)f
+\frac{B^2\beta_3^2}{r^2}\right].
\end{eqnarray}

\section{Superresonance phenomenon} \label{superresonance}

\subsection{Reflection coefficient}

Now, we introduce the metric (\ref{metrinv}) in the relation
Klein Gordon given by
\begin{eqnarray}
 \frac{1}{\sqrt{-g}} \partial_{\mu} ( \sqrt{-g} g^{\mu\nu}
 \partial_{\nu} \phi_{1}) = 0 \label{eq14'}
\end{eqnarray}
Bearing in mind the stationarity and axisymetry spherical metric
(\ref{metrinv}), we can separate the potential as follows:
\begin{eqnarray}
 \phi_1 (t,r,\theta) = R(r)\; e^{i\;(w\;t\;-m\;\theta)} \label{eq33}
\end{eqnarray}
where $m$ is a real constant (azimuthal  number) and $ w $ the
rotational frequency of the acoustic BH. Considering (\ref{metrinv}),
(\ref{eq14'}), (\ref{eq33}), The radial function $R(r)$ satisfies
the linear second order differential equation

\begin{eqnarray}
R''(r)+ P_1(r) \; R'(r) + P_2(r)\;R(r)  =0           \label{eq34}
\end{eqnarray}
where
\begin{eqnarray}
 P_1(r) = \frac{\Gamma_1}{\Gamma_2}  \label{eq34'}
\end{eqnarray}

\begin{eqnarray}
\Gamma_1 &=&  \Big \lbrace c^2_s\; r^2 [ 8c^2_s\;r^2 +
B^2(-1+\gamma)^2]+ 3A^4 \;\gamma(-1+\gamma)^2+
A^2[3B^2(-1+\gamma)^2\; \cr \label{eq34a'} &\times&\gamma + c^2_s\;
r^2(1+\gamma(6+\gamma))] \Big \rbrace\\\label{eq34b'} \Gamma_2 &=&
2r[4c^2_s\; r^2+A^2(-1+\gamma)^2+ B^2(-1+\gamma)^2](c^2_s\; r^2- A^2
\;\gamma)^2
\end{eqnarray}

and
\begin{eqnarray}
 P_2(r) = \frac{\Gamma_3}{\Gamma_4}  \label{eq34''}
\end{eqnarray}
where
\begin{eqnarray}\nonumber
&& \Gamma_3  = r^2\Big[4c^2_s +A^2 (-1+\gamma)^2 + B^2(-1+\gamma)^2
\Big] \Big[ -4c^4_s\; r^2 \;m^2 + A^2\;r^2\;w^2\\ \nonumber &&
(-1+\gamma)^2 \Big]+ 4c^2_s\;[(A^2+B^2)m^2 \;\gamma - i \;
B\;m\;r\;w(1+\gamma)+r^4\;w^2],
\end{eqnarray}
 and
 \begin{eqnarray}
\Gamma_4  = 4 \Big[4c^2_s\; r^2+A^2(-1+\gamma)^2+B^2(-1+\gamma)^2
\Big] \Big(c^2_s\; r^2- A^2 \;\gamma \Big)^2.
\end{eqnarray}

Thus, the problem has reduced to a one dimensional Schr\"{o}dinger problem.
Two further analytical simplifications can be made (\ref{eq34}): first,
We now introduce the tortoise coordinate $r^{\ast}$ by using the following equation
\begin{eqnarray}
\frac{d}{dr^{\ast}}=\left(1-\frac{r_{h}^2}{r^2}\right)\frac{d}{dr},
\end{eqnarray}
which gives the solution
\begin{eqnarray}
r^{\ast} =\Big(1+ \frac{(A^{2} + B^{2})}{4\;c_s^2}(\frac{\gamma -1}{2})^2
\Big)  \Big(r+ \frac{|A|\;\sqrt{\gamma}}{2\;c_{s} }\; \log \Big|
\frac{r\; c_s - A\; \sqrt{\gamma}}{r\; c_s + A\; \sqrt{\gamma}}
\Big| \Big).
\end{eqnarray}

Note that the tortoise coordinate spans the entire real line as
opposed to $r$ which spans only the half-line; the
horizon $r =  \frac{\sqrt{\gamma} \;|A|}{c_{s}}$ maps to $ r^{\ast} \longrightarrow -\infty $, while
$ r \longrightarrow \infty $ corresponds to $ r^{\ast} \longrightarrow +\infty $.
Next, introducing a new radial function $Z(r)\;G(r^{\ast})=  R(r) $.
The equation (\ref{eq34}) becomes:
\begin{eqnarray}
 \frac{d^{2} G(r^{\ast})}{dr^{\ast2}} + Q(r)\; G(r^{\ast}) = 0 \label{eq35}
\end{eqnarray}
with

\begin{eqnarray}\nonumber
Q(r)&=& \frac{1}{\Delta^2}\; \Big(\frac{1}{Z(r)} \frac{d^{2}
Z(r)}{dr^{2}} +P_1\;\frac{1}{Z(r)} \frac{d Z(r)}{dr}+P_2
\Big)\\\nonumber &=& \frac{\frac{(-1 + (A^2 \gamma])}{(cs^2 r^2))^2
}}{16\Big(1+\frac{(A^2+B^2)(-1+\gamma)^2}{4c^2_s}\Big)^2}  \Big
\lbrace  \frac{(144 cs^4 r^2)}{(4 cs^2 r^2 + A^2 (-1 + \gamma])^2 +
B^2 (-1 + \gamma])^2)^2}\\ \nonumber &+&\frac{5}{r^2} -\frac{ (8
cs^2)}{(4 cs^2 r^2 + A^2 (-1 + \gamma])^2 + B^2 (-1 + \gamma])^2)}-
\Big \lbrace       2 (8 cs^2 r^2 + A^2 (-1 \\ \nonumber &+&
\gamma])^2 + B^2 (-1 + \gamma])^2) (cs^2 r^2 (8 cs^2 r^2 + B^2 (-1 +
\gamma])^2) + 3 A^4 (-1 + \gamma])^2\\ \nonumber &\times&  \gamma] +
A^2 (3 B^2 (-1 + \gamma])^2 \gamma] + cs^2 r^2 (1 + \gamma] (6 +
\gamma]))))  \Big \rbrace / \Big \lbrace r^2 (4 cs^2 r^2 +  \\
\nonumber &\times& A^2 (-1 + \gamma])^2 + B^2 (-1 + \gamma])^2)^2
(cs^2 r^2 - A^2 \gamma])   \Big \rbrace +\Big \lbrace 4 r^2 (4 cs^2
+ A^2\\\nonumber &\times&  (-1 + \gamma])^2 +  B^2 (-1 + \gamma])^2)
(-4 cs^4 m^2 r^2 +A^2 r^2 (-1 + \gamma])^2 \omega]^2 +
\\\nonumber && 4 cs^2 ((A^2 + B^2) m^2 \gamma] -     B m r^2 (1 +
\gamma]) \omega] + r^4 \omega]^2))   \Big \rbrace / \Big \lbrace (4
cs^2 r^2 + A^2\\\nonumber &\times& (-1 + \gamma])^2 + B^2 (-1 +
\gamma])^2) (cs^2 r^2 - A^2 \gamma])^2 \Big \rbrace.
\end{eqnarray}

We remark that the main advantage of the new radial equation
(\ref{eq35}) over (\ref{eq34}) is the absence in the former of a
first derivative of the radial function. We analyze differential
equation (\ref{eq35}) in two distinct radial regions near the
horizon, i.e $r^{\ast} \longrightarrow -\infty $ and at asymptote,
i.e $r^{\ast} \longrightarrow +\infty $. In the asymptotic region,
Eq.(\ref{eq35}) can be written approximately as,
\begin{eqnarray}
 \frac{d^{2} G(r^{\ast})}{dr^{\ast2}} +\sqrt{\frac{w^2}{c^2_s+\frac{1}{4}(A^2+B^2)(-1+\gamma)^2}}  \;\;G(r^{\ast}) =0
\end{eqnarray}
of which one solution is
\begin{eqnarray}
G(r^{\ast})&=& \exp
\Big[i\;r^*\;\sqrt{\frac{w^2}{c^2_s+\frac{1}{4}(A^2+B^2)(-1+\gamma)^2}}
\Big]  \cr &+& \mathcal{R}\;\exp \Big [-i\;r^*
\;\sqrt{\frac{w^2}{c^2_s+\frac{1}{4}(A^2+B^2)(-1+\gamma)^2}}\Big].
\label{eq36}
\end{eqnarray}
where $\mathcal{R}$ is the reflection coefficient. In equation
(\ref{eq36}), the first term is the incident wave and the second
term is the reflected wave. The Wronskian of solution (\ref{eq36})
turns out to be
\begin{eqnarray}
W(+\infty) = -2\;i
\sqrt{\frac{w^2}{c^2_s+\frac{1}{4}(A^2+B^2)(-1+\gamma)^2}}
\;(1-|\mathcal{R}|^{2}).  \label{eq37}
\end{eqnarray}
Considering the second field ($r^{\ast} \longrightarrow -\infty $),
the equation (\ref{eq35}) becomes:
\begin{eqnarray}
&& \frac{d^{2} G(r^{\ast})}{dr^{\ast2}}+ \Big \lbrace
4(-2B\;c^2_s\;m+A^2(1+\gamma)\;w)^2   \Big \rbrace / \Big \lbrace
A^2(4c^2_s\; r^2+A^2 (-1+\gamma)^2\cr &+&  B^2(-1+\gamma)^2)(A^2
(1+\gamma)^2+B^2 (-1+\gamma)^2)  \Big \rbrace  G(r^{\ast})=0. \label{eq38}
\end{eqnarray}
The physical solution of equation (\ref{eq38}) is the following.
\begin{eqnarray}
G(r^{\ast})&=& \mathcal{T} \exp\; -ir^* \Big \lbrace
4(-2B\;c^2_s\;m+A^2(1+\gamma)\;w)^2 \Big \rbrace  \Big / \Big
\lbrace A^2(4c^2_s\; r^2\cr \nonumber&+&A^2 (-1+\gamma)^2+ B^2
(-1+\gamma)^2)(A^2 (1+\gamma)^2+B^2 (-1+\gamma)^2) \Big \rbrace.
\\\label{eq40}
\end{eqnarray}
where $\mathcal{T}$ is the transmission coefficient.
 The Wronskian of solution (\ref{eq40}) is following
\begin{eqnarray}
W(-\infty) &=& \Big \lbrace  \Big[ -2\;i \;
4(-2B\;c^2_s\;m+A^2(1+\gamma)\;w)^2\Big]\;\;\mathcal{T}^{2} \Big
\rbrace / \Big \lbrace   A^2\cr &\times&(4c^2_s\; r^2+A^2
(-1+\gamma)^2+B^2 (-1+\gamma)^2) (A^2 (1+\gamma)^2\cr &+&B^2
(-1+\gamma)^2)\Big \rbrace .   \label{eq41}
\end{eqnarray}
Since both equations are actually limiting approximations of the
differential Eq. (\ref{eq35}), which, as we have mentioned, has a
constant Wronskian, it follows that (\cite{j}, \cite{j1}, \cite{j2},
\cite{j4})
\begin{eqnarray}
 W(-\infty) = W(+\infty),
\end{eqnarray}
so that, from (\ref{eq37}) and (\ref{eq41}), we obtain the relation
\begin{eqnarray}
|\mathcal{R} |^{2}= 1- \frac{[\; 1+\gamma - \frac{2B\;m\;c^2_s\; }
{w \;A^2} \;]}{\sqrt{(A^2 (1+\gamma)^2+B^2 (-1+\gamma)^2)}}
\mathcal{T}^{2}. \label{eq42}
\end{eqnarray}
superresonance phenomenon occurs when the norm of the reflected
wave is greater than the norm of the incident wave, that
is, when the reflection coefficient is greater than unity \cite{ines1,ines2}. Hence,
we can observe in eq. (\ref{eq42}) that, for frequencies in the
range
\begin{eqnarray}
0 < w < m \;\bar{\Omega}_H
\end{eqnarray}
with
\begin{eqnarray}
\bar{\Omega}_H = \frac{2 }{1+\gamma}\;\;\Omega_{H},\quad \Omega_{H}=
\frac{B \;c^2_{s}}{A^{2}},
\end{eqnarray}
the reflection coefficient has a magnitude larger than unity whose
imply the amplification relation of the ingoing sound wave near
horizon. with this condition we can extract the energy of the system
\cite{j4}. It is noteworthy that $\gamma
> 1$ makes the interval to have superradiance smaller. Here $m$ is
the azimuthal mode number and $ \bar{\Omega}_H$ is the angular
velocity of the usual Kerr-like acoustic BH. the angular velocity of
the usual Kerr-like acoustic BH depends of the pressure (parameters
$n$ and $k$).  Thus, we show that the presence of the the
neo-newtonian parameter $\gamma$ modifies the quantity of removed
energy of the acoustic BH and that is either possible to accentuate
or attenuate the amplification of the removed energy of the
acoustic. The effect of superresonance can be eliminated when $
\gamma = (\frac{2 m \Omega_{H}}{\omega} -1 $).

\subsection{Numerical results}\label{numerical}

\begin{figure}[htbp]
\begin{center}
\includegraphics[width=14cm, height=8cm]{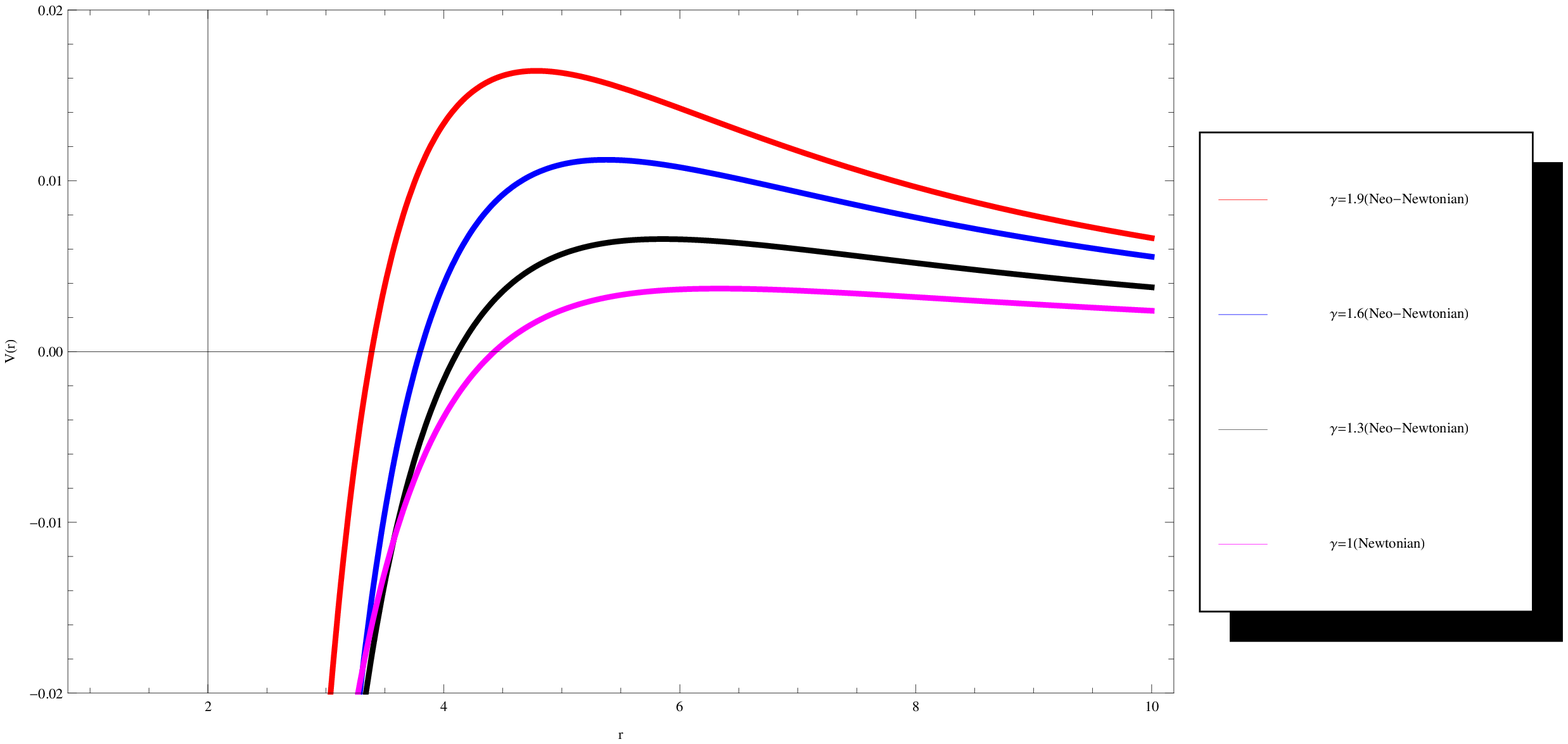}
\end{center}
\caption{\small{The graphs illustrating the  evolution of the
potential $V(r)$ for each value of $\gamma$. The function  is
plotted for $ A=B=c_s=m=1 $.}}\label{fig}
\end{figure}

In order to confirm the effect of pressure on the phenomenon of
superresonance, it is necessary to extract the potential energy. The
objective in this section, would be to see if the pressure changes
the potential energy In this way we obtain the following
differential equation for the radial function $R(r)$ (\ref{eq34}) as
\begin{eqnarray}
\label{EQKG}
&&\left[\Big(1+\frac{\beta_4}{r^2}\Big)\omega^2-\frac{2B\beta_3 m\omega}{r^2}-\frac{m^2f}{r^2}\right]
\frac{(1+\beta_2)R(r)}{(-g){\cal Q}}
+\frac{1}{r\sqrt{-g}}\frac{d}{dr}\left[r\sqrt{-g}(1+\beta_2)^{-1}{\cal Q}\frac{d}{dr}\right]R(r)=0.
\end{eqnarray}
We can rewrite the equation (\ref{EQKG}) as
\begin{eqnarray}
\label{EQKGbtz}
&&\left[\Big(1+\frac{\beta_4}{r^2}\Big)\omega^2-\frac{2B\beta_3 m\omega}{r^2}-\frac{m^2f}{r^2}\right]R(r)
+\frac{{\cal F}(r)}{r}\frac{d}{dr}\left[r{\cal F}(r)\frac{d}{dr}\right]R(r)=0,
\end{eqnarray}
where $ {\cal F}(r)=\sqrt{-g}(1+\beta_2)^{-1}{\cal Q}(r) $.
At this point we introduce the coordinated $\varrho$ using the following equation~\cite{Gamboa}
\begin{eqnarray}
\frac{d}{d{\varrho}}={\cal F}(r)\frac{d}{dr},
\end{eqnarray}
and now introducing the new radial function
$G(\varrho)=r^{1/2}R(r)$, we can obtain the following modified
radial equation
we get to a new radial equation obtained from (\ref{EQKGbtz}) that
reads
\begin{eqnarray}
\label{EG}
\frac{d^2G(\varrho)}{d\varrho}+\left\{\left[\left(1+\frac{\beta_4}{r^2}\right)^{1/2}\omega
-\frac{B\beta_3 m}{r^2}\left(1+\frac{\beta_4}{r^2}\right)^{-1/2} \right]^2-V(r)\right\}G(\varrho)=0,
\end{eqnarray}
and the potential $V(r)$ is given by~\cite{Dolan, ABP2012-1}
\begin{eqnarray}
\label{potv}
V(r)=\frac{{\cal F}(r)}{4r^2}\left[\frac{4m^2f(r)}{{\cal F}(r)}-{\cal F}(r)
+\frac{4B^2\beta_3^2m^2}{r^2{\cal F}(r)}\left(1+\frac{\beta_4}{r^2}\right)^{-1}+2r\frac{d{\cal F}(r)}{dr}\right].
\end{eqnarray}
We note that the equation (\ref{potv}) does not satisfy the
asymptotic behavior $V(r)\rightarrow 0 $ as $ r\rightarrow\infty $.
We can see from  Fig.(\ref{fig}) when the $\gamma$ parameter becomes
larger, the potential energy increases, which allows us to say that
the presence of pressure ascents the phenomenon of superradiance.

One important aspect of the results reported here is that, perhaps,
the specific form of the neo-Newtonian equations, retaining the
essential features of the relativistic problem, may depend on the
symmetries of the problem, as a comparison with the cosmological
case suggests. We have constructed acoustic black hole
configurations in the context of the neo-Newtonian hydrodynamics.
The presence of the parameter $\gamma$ in the effective metric of
acoustic BHs in the framework of the neo-Newtonian theory is the
main feature of our results. This parameter is always larger than
$1$ if $k$ and $n$ have the same sign, that is, if the fluid have
the expected positive square speed of sound. Hence, essentially, the
velocity on the horizon is diminished by the presence of the
pressure. It is worth noting that when $\gamma$ is equal to $1$, the
results of Newtonian theory are recovered \cite{M,fcreation}, which
leads us to rewrite $\gamma$ as follows
\begin{eqnarray}
 \gamma_{neonewtonian}= \gamma_{newtonian} +  \frac{k\,n\,\rho_0^{n-1} }{c^2}.
\end{eqnarray}
The supplementary term is the contribution of the pressure in
neo-Newtonian theory.

\section{Conclusion}\label{conclusion}

In this paper we shown that the presence of the pressure modify the
quantity of removed energy of the acoustic BH and that, it is
possible to accentuate or to attenuate the amplification of the
removed energy of the acoustic BH and still exists the possibility
to cancel the superradiance effect i.e the reflec- tion coefficient
is equal to unity, when $ \gamma = (\frac{2 m \Omega_{H}}{\omega} -1
$) where $\Omega_{H}$ is the angular velocity of the acoustic black
hole. Furthermore, the interval of frequencies becomes narrower in
the present scenario. As a consequence, the tuning of the
neo-newtonian parameter $\gamma$ changes the rate of loss of mass of
the acoustic BH.
\par \bigskip

{\bf Acknowledgement}:  Authors thank Prof. J. C. Fabris and Prof. H. E. S. Velten
 for useful comments and discussions. Ines G. Salako thanks IMSP  for hospitality during the elaboration of this
work.

.


\end{document}